\documentclass[conference]{IEEEtran}
\usepackage{graphicx}
\usepackage{amssymb}
\def\BibTeX{{\rm B\kern-.05em{\sc i\kern-.025em b}\kern-.08em
		T\kern-.1667em\lower.7ex\hbox{E}\kern-.125emX}}
\usepackage{xcolor}
\usepackage{cite}
\usepackage{amsmath,amssymb,amsfonts}
\usepackage{algorithmic}
\usepackage{textcomp}

\newcommand{\Ugtrless}{%
	\mathrel{\kern0pt\mathop{\gtrless}\limits^{1}_{0}}%
}

\begin{document}
	\title{Toward a Wired Ad Hoc Nanonetwork}
	\author{\IEEEauthorblockN{Oussama Abderrahmane Dambri and Soumaya Cherkaoui}
		\IEEEauthorblockA{INTERLAB, Engineering Faculty, Universit\'e de Sherbrooke, Canada
			\\ 
			Email: \{abderrahmane.oussama.dambri, soumaya.cherkaoui\}@usherbrooke.ca}
	}
	\maketitle
	\begin{abstract}
		Nanomachines promise to enable new medical applications, including drug delivery and real time chemical reactions' detection inside the human body. Such complex tasks need cooperation between nanomachines using a communication network. Wireless Ad hoc networks, using molecular or electromagnetic-based communication have been proposed in the literature to create flexible nanonetworks between nanomachines. In this paper, we propose a Wired Ad hoc NanoNETwork (WANNET) model  design using actin-based nano-communication. In the proposed model, actin filaments self-assembly and disassembly is used to create  flexible nanowires between nanomachines, and  electrons are used as carriers of information. We give a general overview of the application layer, Medium Access Control (MAC) layer and a physical layer of the model.  We also detail the analytical model of the physical layer using actin nanowire equivalent circuits, and we present an estimation of the circuit component's values. Numerical results of the derived model are provided in terms of attenuation, phase and delay as a function of the frequency and distances between nanomachines. The maximum throughput of the actin-based nanowire is also provided, and a comparison between the maximum throughput of the  proposed WANNET, vs other proposed approaches is presented. The obtained results prove that the proposed wired ad hoc nanonetwork can give a very high achievable throughput with a smaller delay compared to other proposed wireless molecular communication networks.
	\end{abstract}

	\begin{IEEEkeywords}
		Actin; Nano-Communication; Wired; Ad Hoc; Nanonetworks; Equivalent Circuit; Attenuation; Delay
	\end{IEEEkeywords}
	\section{Introduction}
In recent decades, nanotechnology's progress with the newly emerged nanomachines gave means to promising applications in medical and pharmaceutical fields, such as monitoring and drug delivery [1]. The limited capacity of the nanomachines and the complexity of the medical applications paved the way to a big interest in nanonetworks design, which helps overcome the nanomachines energy and the processing capacity limitations. Nanonetworks is a paradigm that adapts classical communication paradigms to meet the requirement of the nanosystems [2]. Several studies proposed adaptations to the classical wireless and mobile ad hoc networks to be used in nanosystems using either electromagnetic or molecular communications [3]-[5]. 

A downscaled version of the traditional wireless ad hoc is proposed in [3] using carbon nanotube-based communication. The authors investigated the hardware components needed for such networks at nanoscale, presented the networking issues and discussed the advantages of applying ad hoc nanosystems. However, the electromagnetic waves used in carbon nanotube-based communication oscillate at terahertz frequencies, and the design of transceivers that can capture the band peculiarities is one of its main challenges. The safety to use terahertz frequencies inside the human body for medical applications is another problem of electromagnetic communication at nanoscale, which needs further research. Molecular communication is a bioinspired paradigm that uses molecules as information carriers instead of electromagnetic waves. Molecular communication is used by nature and can be leveraged safely inside the human body. In [4], the authors proposed a mobile ad hoc nanonetwork with collision-based molecular communication. This Mobile Ad hoc Molecular NETwork (MAMNET) employs infectious disease spreading principles using the electrochemical collision between mobile nanomachines to transfer information. Another MAMNET system is proposed in [5], by using F\"orster Resonance Energy Transfer (FRET-MAMNET), which is a nonradiative excited state energy transfer phenomenon, to send information from a donor to a nearby acceptor using fluorophores. However, the achievable throughput of both proposed methods drastically decreases with the increase in the distance between nanomachines.

In our recent work [6], we proposed self-assembled actin-based nano-communication, which uses actin filaments as a conductive nanowire to transmit electrical information and uses magnetic field for the actin direction. Actin is a bi-globular protein that exists in all human cells, and its ability to self-assemble and disassemble allows to create a flexible wired nanonetwork between mobile nanomachines. We also analyzed the stability of the nanowire with stochastic simulations, using our developed framework and we calculated the error probability.

In this paper, we take a first step toward designing a wired ad hoc nanonetwork by using the actin-based nano-communication method proposed in our recent work [6]. In the proposed model, the actin filaments self-assembly create conductive nanowires between nanomachines, and uses electrons as carriers of information. We demonstrate that once a link is established between an emitter and a receiver, the achievable throughput of the proposed Wired Ad hoc NanoNETwork (WANNET) is independent of the distance between nanomachines. The natural assembly and disassembly of actin filaments can be controlled via enzymes without the need of any infrastructure to create a flexible nanowire, which allows us, for the first time in the literature, to design wired ad hoc networks at the nanoscale. In contrast to the previous work, the proposed system in this paper does not use magnetic field to direct actin filaments, its direction is left random so that the ad hoc system will be flexible and without any infrastructure. A proposed nanomachine is designed to trigger the formation of an actin nanowire when detecting a desired substance, which links the nanomachine with one of the surrounding neighbors. The transmitter can use the piezoelectric property of some proteins and DNA to transfer mechanical energy into electricity and send it through the constructed actin nanowire to the receiver. The receiver then triggers the formation of another actin nanowire and repeats the transmission cycle. In this paper, We also propose the OSI model-based communication layers between two nanomachines of the WANNET, and we briefly explain the role of each layer. Moreover, we derive an analytical model of the physical layer represented by the actin filaments, and we use this model to obtain the numerical analysis in order to evaluate the physical layer's performance.

The rest of the paper is organized as follows. In section II, we derive the analytical model of the physical layer represented by the actin filaments, by considering the actin monomers as an equivalent RLC circuit, and we calculate its maximum throughput. In section III, we explain the proposed wired ad hoc system and we detail the other two layers namely, application layer and MAC layer. We provide numerical results in section IV, by using the derived model, and we evaluate the performance of the physical layer in terms of attenuation, phase and delay as a function of the frequency and distances between nanomachines. We also provide an approximation of the maximum throughput of the proposed WANNET, and we compare it with the maximum throughput of FRET-MAMNET proposed in literature [5]. Finally, section V concludes the paper and presents future work.
 
	\section{Physical Layer Model}
	 
The physical layer, which is the main subject of this paper, uses the assembled actin nanowire. Actin is a bi-globular protein with self-assembly ability, which is controlled by specific enzymes. We assume that the distance between nanomachines does not exceed few tens of micrometers and that the radius of the nanomachines is not too small compared to the distance between them. These assumptions increase the probabilty of finding a neighbor nanomachine by the assembled nanowire, because the actin self-assembly direction is random. Moreover, actin has electrical properties that allow it to play the role of a conductive nanowire [7]. Several experimental studies show that actin filaments can be used as electrical conductive wires [8], [9]. 

The objective of modeling the ad hoc system's physical layer is to study the attenuation, phase and the delay as functions of the frequency and the channel distance. The actin monomers are modeled in the literature as an equivalent circuit based on the analogy of transmission lines with resistive, inductive and capacitive components [10], [11]. Each infinitesimal element of a transmission line length can be represented as a resistive and inductive components linked in series, and a capacitance with another resistive components linked in parallel. Due to the constant current in all components, transmission lines are usually modeled as infinite RLC equivalent circuits linked in series. In our proposed physical layer model, the infinitesimal elements of the actin filament are the actin monomers, and each monomer is modeled as a series RLC equivalent circuit as shown in Fig. 1.

	 \begin{figure}
	\centering
	\includegraphics[width=.8\linewidth]{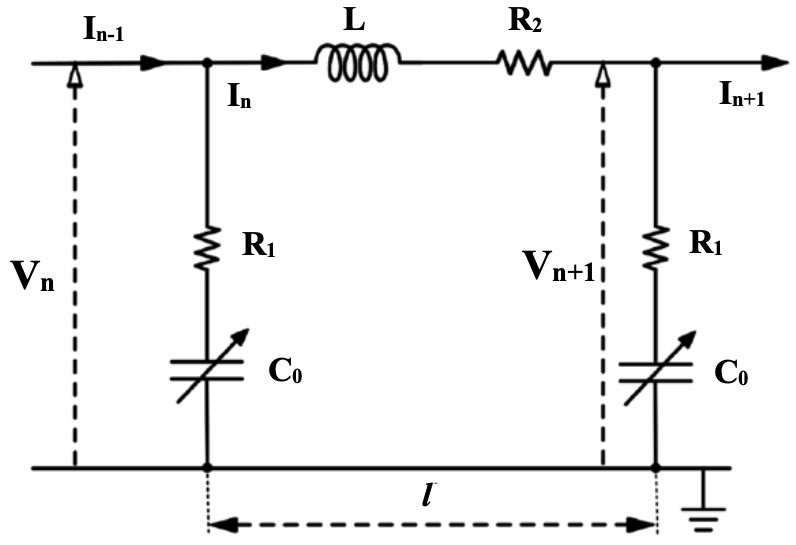}
	\caption{Effective circuits diagram for the n$^{th}$ monomer of an actin filament.}
	\label{fig:1}
\end{figure}

\begin{figure}
	\centering
	\includegraphics[width=.8\linewidth]{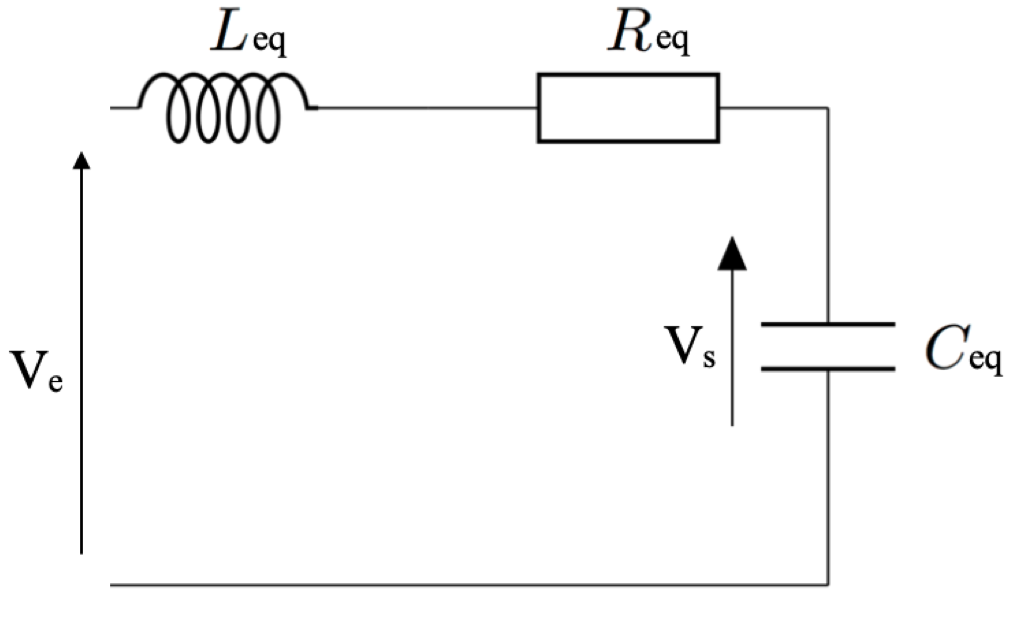}
	\caption{Effective circuits diagram for the total actin filament.}
	\label{fig:2}
\end{figure} 
	 
	  \subsection{Circuit's Components}
	  
	  Before explaining each of the component's physical significance, we identify the Bjerrum length $\lambda_B$ as the distance where the electrostatic attractions of the actin filaments charges are stronger than the thermal fluctuations. The value of a Bjerrum length is found to be $\lambda_B$ = 7.13 $\times$ $10^{-10}$ m for a temperature of 293 K [10]. The negative charge of proteins constructing the actin filament and the positive countrions around the filament create a $\sim 1$ Bjerrum radius depletion area. The depletion area is modeled as a capacitance in the equivalent circuit shown in Fig. 1,  and it can be calculated as:
  
     	\begin{equation}
  C_0= \frac{2\pi \varepsilon l}{ln \left(  \frac{r_{actin} + \lambda_B}{r_{actin}}\right)  }. 
  		\label{eq:1}
 		\end{equation} where $\varepsilon$ is the permittivity, $r_{actin}$ is the radius of actin filaments and $l$ is the length of an actin monomer and can be approximated to $l$$\simeq$ 5.4 nm. The estimated capacitance per monomer can be approximated to  $C_0 \simeq $ 96 $\times$ 10$^{-6}$ pF.
	 
The movement of the charges in the helical shape of actin filaments induces a magnetic field, which is modeled as an inductance. The effective inductance for the actin filament in a solution can be calculated:

     	\begin{equation}
	L= \frac{\mu H^2A}{l}.
		\label{eq:2}
		\end{equation} where $\mu$ is the magnetic permeability, $A$ is the cross-sectional area $A=\pi (r_{actin}+\lambda_B)^2$ and $H$ is the number of the helical turns in the actin filament. The estimated inductance per monomer is $L \simeq $ 1.7 pH. 
	
	The authors in [10] calculated the resistance of actin monomers based on conductivities of ions in a salt solution as:
	
     	\begin{equation}
		R= \frac{\rho ln ((r_{actin} + \lambda_B)/r_{actin})}{2\pi l}. 
		\label{eq:3}
		\end{equation} where $\rho$ is the resistivity. The estimated resistance per monomer is $R \simeq $ 6.11 M$\Omega$.
	
	To study the electrical characteristics of an entire filament that contains $n$ monomers, and by assuming that the values of all the components for each monomer are equal, we sum the $n$ circuits in an equivalent circuit as presented in Fig. 2. In [10], the authors estimated an average of 370 monomers per $\mu$m. By summing 370 RLC circuits linked in series we got an effective resistance, inductance and capacitance for a 1 $\mu$m of the actin filaments as follows:

		 	\begin{flushleft} 
		 		\centering
				$C_{eq}$ = 0.02  $\times$ 10$^{-12}$ F

				$L_{eq}$= 340  $\times$ 10$^{-12}$  H

				$R_{eq}$= 1.2 $\times$  10$^{9} \Omega$
			\end{flushleft} 

    \subsection{Circuit Analysis}
  
From the equivalent RLC circuit presented in Fig. 2, we derive the transfer function as follows:

	\begin{equation} \label{eq:4}
		\begin{split}
	T= \frac{V_s}{V_e}& = \frac{\frac{1}{sC_{eq}}}{sL_{eq} + R_{eq} + \frac{1}{sC_{eq}}}  \\
	&= \frac{1}{L_{eq} C_{eq}}   \cdot \frac{1}{s^2 + s\frac{R_{eq} }{L_{eq}} + \frac{1}{L_{eq} C_{eq}}},
		\end{split}
	\end{equation} Where $s$ represents the frequency. To lighten the expression, we write $C_{eq}$, $L_{eq}$ and $R_{eq}$ as $C$, $L$ and $R$ respectively. The denominator in (4) gives two solutions:

     	\begin{equation}
	\text{\footnotesize $	p_{1,2}= -\frac{R}{2L} \pm \sqrt{\left(\frac{R}{2L} \right)^2- \frac{1}{LC}}, $}
		\label{eq:5}
		\end{equation}

Equation (5) includes 3 different cases depending on the comparison between the two expressions ($R/2L$)$^2$ and $1/$($LC$). In our study, due to the very high actin filament resistance, we are  in the case where ($R/2L$)$^2$ is bigger than $1/$($LC$). The attenuation $A(s)$ is then calculated in dB as:

\begin{equation}
 A(s)|_{dB} = 20 log \frac{1}{LC} - 20 log \sqrt{s^2+p_{1}^2} - 20 log \sqrt{s^2+p_{2}^2}. 
	\label{eq:6}
\end{equation}

The phase $\varphi (s)$ of the derived transfer function is calculated:

\begin{equation}
	\varphi (s) = tan^{-1} \left(\ \frac{1}{LC} \cdot   \frac{1}{(s-p_1)(s-p_2)} \right).
	\label{eq:7}
\end{equation}

The delay $\tau (s)$ can be obtained by calculating the derivative of the phase as:

\begin{equation}
	\tau (s) = - \frac{d \varphi (s)}{ds}.
	\label{eq:8}
\end{equation}

  \subsection{Maximum Throughput}
  
 The electrons are the carriers of information in our proposed actin-based nano-communication method. In order to calculate the maximum throughput that the proposed nanwire can provide, we need to approximate the charge capacity of the actine filaments and the speed at which the charges move through it. The charge capacity of each actin monomer is $\sim 4 e$ [10], and by assuming 370 monomer/$\mu$m of an actin filament, we can deduce that the charge capacity of an actin filament with 1$\mu$m long is approximately 1480 $e$. The velocity expression of charges propagation along actin filaments in units of $m s^{-1}$ can be written as [11]: 
 
 \begin{equation}
 v(t)= \frac { \beta}{ \alpha} \bigg (1+ \frac {1}{24} \frac {d\eta(\tau)}{d\tau} \bigg |_{\tau = t/(24 \alpha)} \bigg ),
 \label{eq:9}
 \end{equation} where 
 
\begin{equation}
\begin{aligned}
\frac {d\eta}{\tau} = 4\Omega^2  \frac {\exp \Big (- \frac {4\tau \mu_2 }{3}\Big )}{1 + \frac {4\mu_1 \Omega^2}{5\mu_2}\Big (1-\exp \Big (- \frac {4\tau \mu_2}{3} \Big )\Big )}, \\
\alpha = \frac {R}{L} + CR > 0, \qquad  \qquad \qquad \beta = 2l.
\end{aligned}
\label{eq:10}
\end{equation} where $\Omega$ = 2.3810, which is dependent on the voltage at the input. $\tau$ is the pulse duration of the electrical signal. $\mu_1 = \frac{6R_1}{R}$,  $\mu_2 = \frac{24R_2}{R}$, where $R_1$ and $R_2$ are the longitudinal and radial ionic flow resistances consecutively.
 
 The results in [11] shows that the velocity of charges propagation along the actin filament starts with 0.03 $m s^{-1}$ and quickly decreases because of the filament's high resistance until it stops in $t$=60$\mu$s. The maximum throughput $T_M (t)$ of the proposed actin-based nano-communication physical layer is the speed of charges propagation multiplied by the total charge capacity of 1$\mu$m long filament, where one electron represents 1 bit of information and we can write:

  \begin{equation}
  	T_M (t) = v(t) \times \psi_{tot}.
  	\label{eq:11}
  \end{equation} where $v(t)$ is the velocity of charges propagation in (4), and $\psi_{tot}$ is the total charge capacity of a 1$\mu$m long actin filament. 
  
  \section{A Wired Ad Hoc Nanonetwork}
  
  An ad hoc network consists of nodes that connect with each other wirelessly, without the need of a mediating infrastructure. Each node can establish a link with its neighbor and communicate with it directly, playing a role to forward data for other nodes [12]. In the literature, ad hoc nanonetworks are proposed by using electromagnetic waves [3] or molecules [4], [5] as a link between nanomachines. However, the proposed nanonetworks based on electromagnetic links still do not addresses some of the challenges related to the peculiarities of the Terahertz band [2], and those based on molecules have a  weak achievable throughput and a high delay [13]-[15].
  
  In this paper, we propose a new method to establish wired links between N nanomachines by using the actin-based method proposed in our recent work [6]. The ability of actin proteins to assemble, create conductive filaments, and then disassemble, allows to use these actin proteins as links between static or mobile nanomachines in a wired ad hoc nanonetwork. Fig. 3 illustrates the proposed WANNET. In the figure, there are 13 nanomachines close to each other. The actin monomers assembly can be triggered when the n$^{th}$ nanomachine receives the desired substance presented as a green sphere by using the chemical cascade in the cells to secrete the assembly enzymes.This in turn, creates a nanowire that connects with a neighbor randomly. After establishing a link, the sender uses the piezoelectric property of some proteins and DNA to send the information using electrons [16]. The received electrons can trigger the creation of a nanowire in the (n+1)$^{th}$ nanomachine that connects with another neighbor randomly by using the same chemical cascade and so on. The cycle of triggering the nanowire creation to send electrons that in turn, triggers another nanowire creation, can help spreading information in all the nanonetwork, until a gateway is reached.

  \subsection{Nanomachines}
  
  The nanomachines in our proposed system are bioengineered cells with DNA and ribosomes, which are already designed and presented in literature as in [17]. The DNA and ribosomes inside the nanomachines enable them to construct proteins and secrete the enzymes that control the actin nanowire formation. As explained earlier, the assumptions about the distance between nanomachines and their radius is due to the fact that the actin self-assembly direction is random. Therefore, small distances between nanomachines and a bigger radius increase the possibility of the random nanowire to find a neighbor. When a nanomachine detects a desired substance to be analyzed, processes take place allowing neighboring nanomachines to communicate information using the constructed nanowire. The nanowire link enables spreading information in all the nanonetwork until reaching the closest gateway. The  proposed WANNET could, ultimately, be used inside the human body for monitoring, disease diagnosis and for real time chemical reactions detection. As illustrated in Fig. 4, the communication processes between nanomachines in the proposed WANNET are performed by three layers:  the physical layer, which we modeled in the previous section, the application layer and the MAC layer.
  
    \begin{figure}
  	\centering
  	\includegraphics[width=\linewidth]{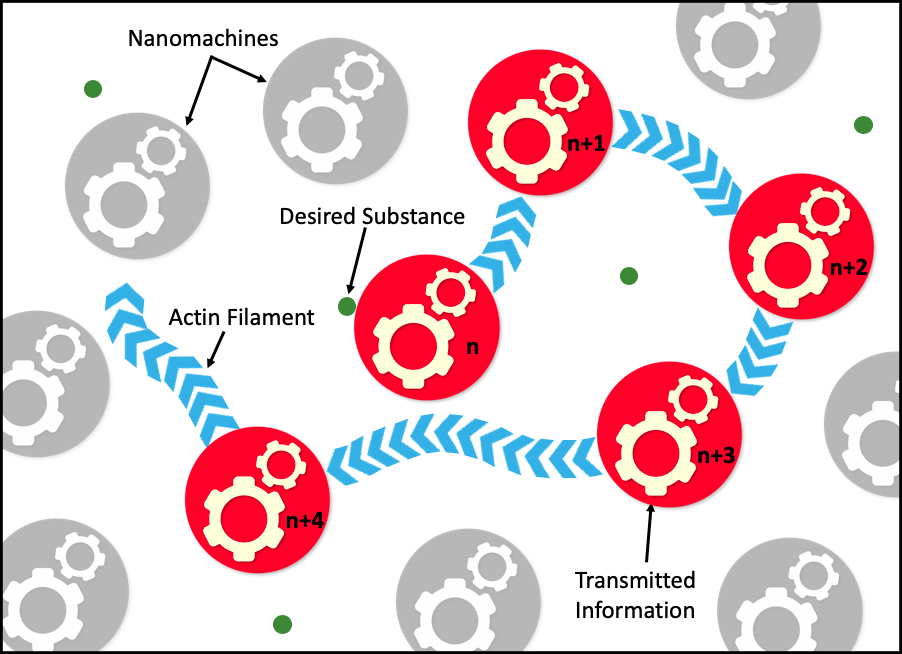}
  	\caption{Information spreading in the proposed wired ad hoc nanonetwork.}
  	\label{fig:3}
  \end{figure}
  
  \begin{figure}
  	\centering
  	\includegraphics[width=\linewidth]{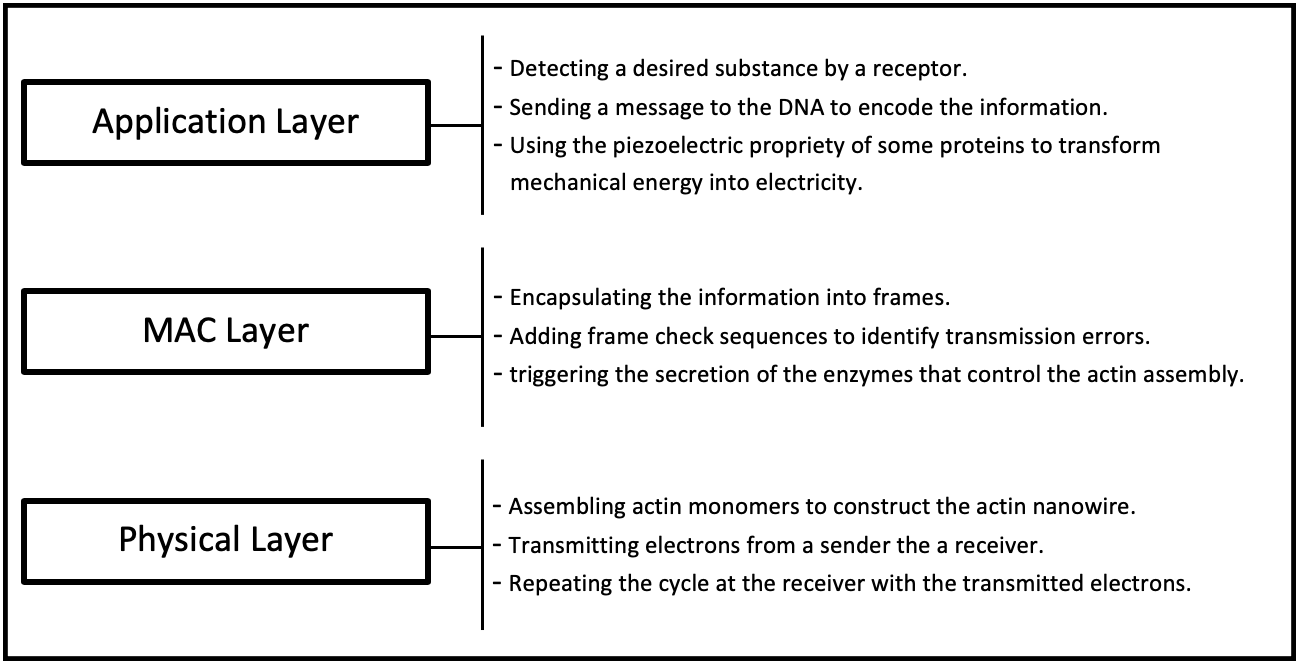}
  	\caption{An OSI Model for the communication layers between two nanomachines.}
  	\label{fig:4}
  \end{figure}
  
  \subsection{Application Layer}
  
  The application layer of the proposed WANNET performs the information encoding. One way to encode the information is that specific receptors at the nanomachine membrane detect the desired substance to be analyzed and send a message to the DNA. DNA encodes the information about the detected substance by using the piezoelectric property of some proteins [16]. These proteins transform mechanical energy into electricity, which is transmitted by using the actin nanowire. The information can be modulated using the dynamic pattern frequency spectrum of the used piezoelectric proteins.
  
  \subsection{Medium Access Control Layer (MAC)}
  
  The MAC layer provides a control over the input of the physical layer by encapsulating the encoded information into frames. It also triggers the secretion of the enzymes that control the actin nanowire formation. In order to identify transmission errors, the MAC layer can also add some frame check sequences. A detailed protocol for the MAC layer will be proposed in future work.

   \section{Numerical Results}

In this section, we present the numerical results of the system's physical layer in terms of attenuation, phase and delay as a function of the frequency and distances between nanomachines, as well as the maximum throughput. We considered a frequency range [0Hz-700Hz] and [0Hz-900Hz] to ease the comparison with the studies that use the same range for molecular communications. Moreover, higher frequencies increase the attenuation of the actin filament, therefore, we expect that the chosen frequency range can give better results than higher frequency ranges. We use MATLAB to obtain the numerical results, and we use the parameters explained above to simulate the electrical characteristics of the system's physical layer. 

   \subsection{Attenuation}

Fig. 5 shows the attenuation of the system model, which is calculated with Eq. (6) for different actin filaments lengths ($d$). The attenuation starts with -364 dB for 0 Hz and $d$=10$\mu$m, and increases with the increase in frequency until it reaches -525 dB for 700 Hz. The high attenuation of the actin filament is due to its very high resistance calculated in Eq. (3). The connection of $n$ circuits representing $n$ monomers to get the actin filament effective circuit, makes the connected capacitances in parallel, and the impedances in series. The increase in attenuation with the frequency increase can be explained by the fact that impedances in series favor low frequencies. Moreover, capacitances in parallel favor the passing of high frequencies and take them away from the output, which makes them play the same role as the series impedances. We can also notice that the attenuation increases with the filament's length increase, for frequencies lower than 400 Hz. The explanation of this increase is that the more the length of actin filaments, the more monomers are needed, which increases the number of impedances in series.

\begin{figure}
	\centering
	\includegraphics[width=.95\linewidth]{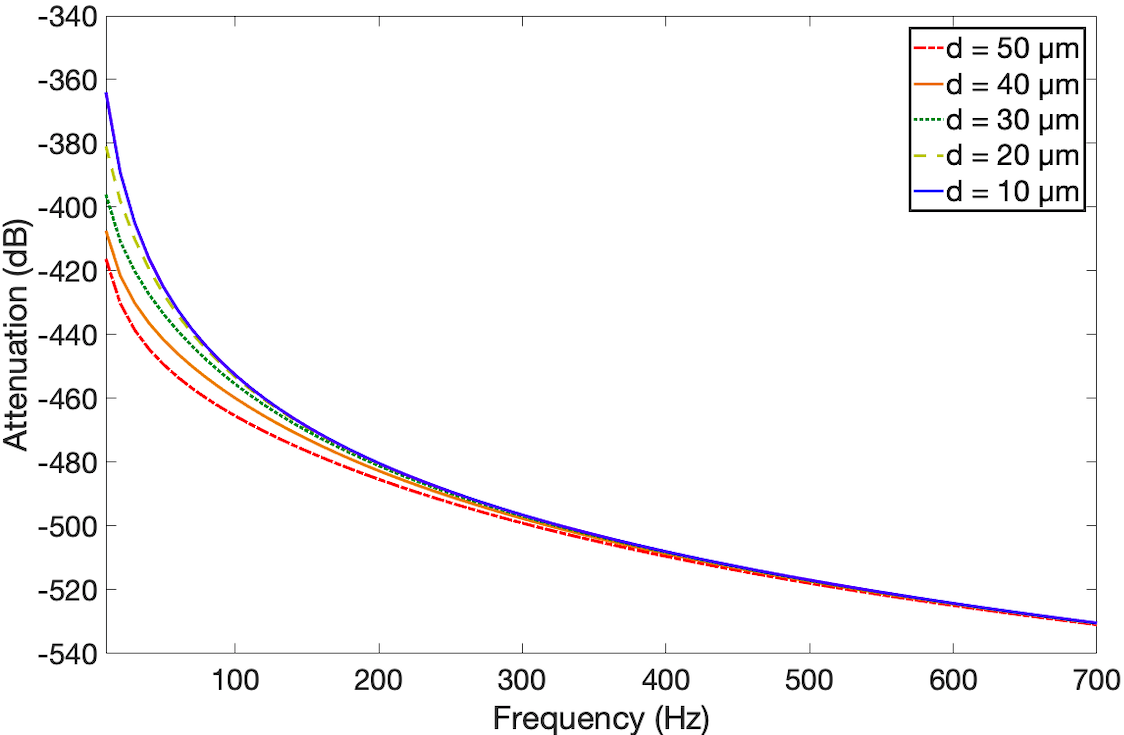}
	\caption{Attenuation for the actin filament model.}
	\label{fig:5}
\end{figure}  

\begin{figure}
	\centering
	\includegraphics[width=.95\linewidth]{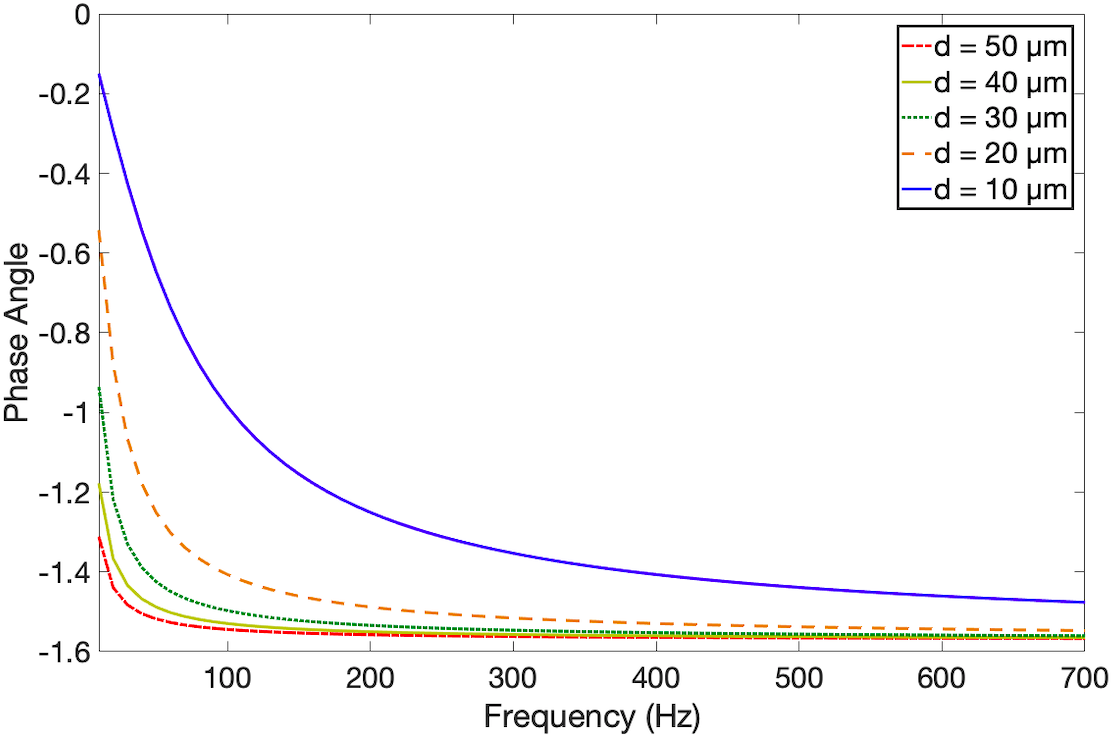}
	\caption{Phase for the actin filament model.}
	\label{fig:6}
\end{figure}

\begin{figure}
	\centering
	\includegraphics[width=.93\linewidth]{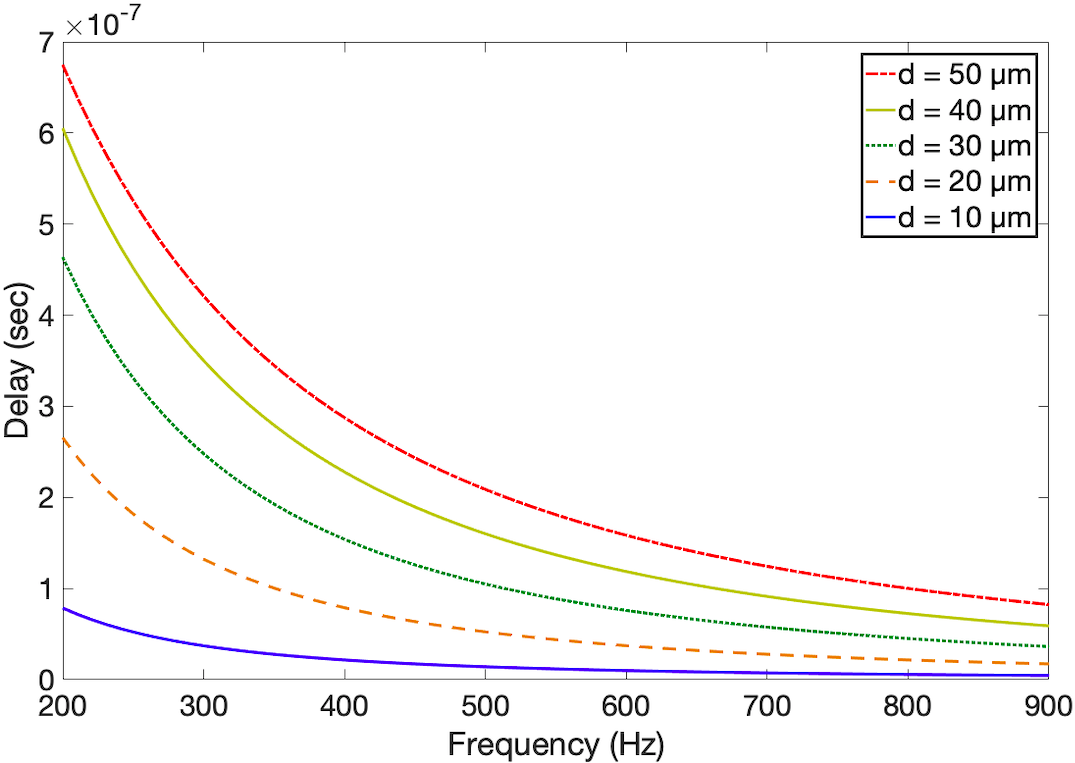}
	\caption{Delay for the actin filament model.}
	\label{fig:7}
\end{figure} 

   \subsection{Phase}
   
Fig. 6 presents the phase of the system model as a function of the frequency, which is calculated with Eq. (7) for different actin filaments lengths. The phase starts with -0.16 degree for 0 Hz and $d$=10$\mu$m. The increase in frequency widen the phase to reach -1.42 degree for 700 Hz. The phase increase is due to the increase in signal attenuation; the more the signal is attenuated the more the phase between the input and the output is increased. The phase also increases with the increase in actin filament's length. The explanation of this increase is that bigger distances crossed by the signal, increase its attenuation as we saw in Fig. 5, and the increase in the signal attenuation also increases  its phase.

   \subsection{Delay}
   
   Fig. 7 presents the delay of the system model as a function of the frequency, which is calculated with Eq. (8) for different actin filaments lengths. The delay starts with 8 $\mu$s for 0 Hz and $d$=10$\mu$m. The increase in frequency decreases the delay to reach 1 $\mu$s for 900 Hz. It is obvious that when frequency increases, the speed of the signal also increases, which decreases the delay. We can also notice that the delay increases with longer actin filaments. The explanation of this increase is that in longer filaments, the signal takes more time to reach the destination,  which increases the delay to reach 68 $\mu$s with $d$=50$\mu$m.

    \begin{figure}
   	\centering
   	\includegraphics[width=.9\linewidth]{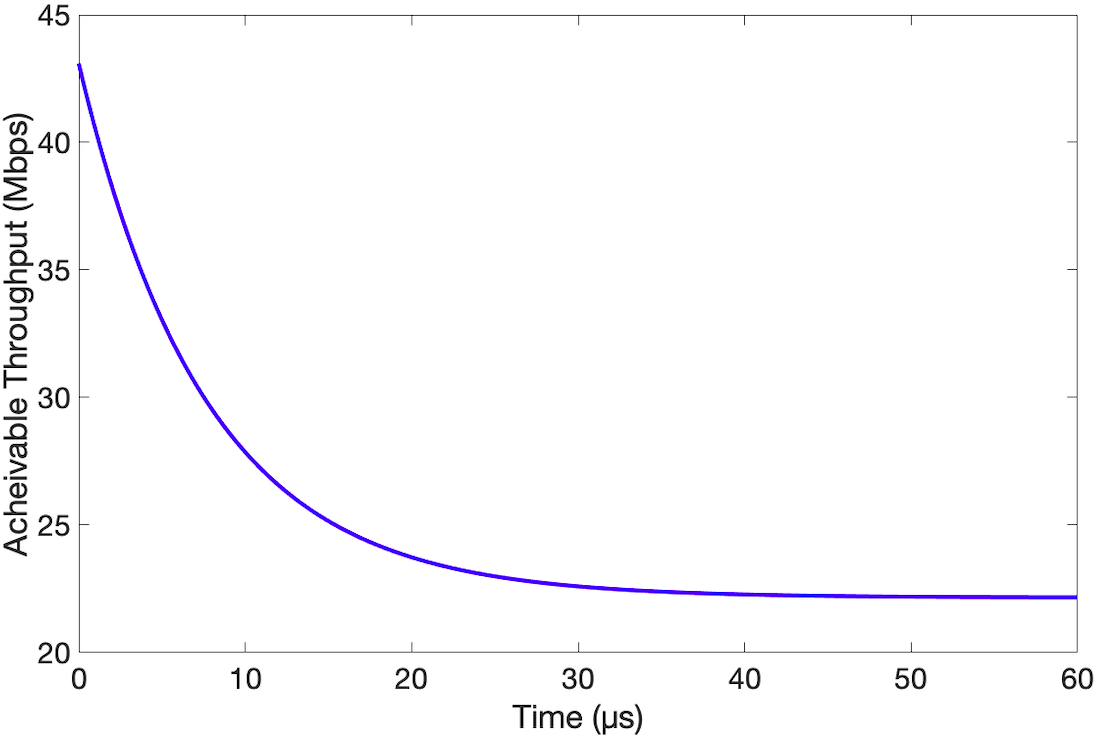}
   	\caption{Maximum throughput approximation for the actin nanowire.}
   	\label{fig:8}
   \end{figure} 

   \subsection{Maximum Throughput}	
	 
	 In Fig. 8, we present the numerical results of the proposed physical layer's maximum throughput, which is calculated with Eq. (11). Charge capacity of 1 $\mu$m actin filament is 1480 $e$, and the speed of the charge propagation is 30,000 $\mu$m/s [11]. If we assume that one electron represents 1 bit, then the maximum throughput of 1 $\mu$m actin filament is 44,4 Mbps as shown in Fig. 8. However, we can notice that the maximum throughput decreases rapidly with time, to reach 23 Mbps in 30 $\mu$s. The explanation of this decrease is that the speed of the charge propagation decreases with time due to the very high resistance of the actin filament as proven experimentally in [13]. Despite this decrease, the throughput still very high and can be used by nanomachines to propagate the information through the proposed WANNET. The best throughput results of the proposed FRET-MAMNET in [5] did not exceed 5.5 Kbps. Because of the huge difference between the results in [5] and the results of our proposed model, we need to plot them logarithmically so that the comparison becomes clear as shown in Fig. 9.

	 \section{Conclusion and Future Work}
	 
	The decentralized nature of ad hoc nanonetworks makes them suitable for a variety of medical and pharmaceutical applications. However, wireless ad hoc nanoetworks proposed in literature either suffer from scattering loss, and molecular absorption or they provide weak throughput with high delay. In this paper, we presented a first step toward designing a Wired Ad hoc NanoNETwork (WANNET) by using actin-based nano-communication. We proposed the OSI model-based communication layers between two nanomachines of the WANNET, and briefly explained the role of each layer. We also derived an analytical model of the physical layer by considering the actin filament as an equivalent RLC circuit. Moreover, we evaluated the performance of the physical layer in terms of attenuation, phase and delay as functions of the frequency and distances between nanomachines. Despite of the fact that the attenuation of the actin filament is very high because of its high resistance, it still provides a very high maximum throughput with smaller delay, compared to the methods that use molecular communication.

	\begin{figure}
		\centering
		\includegraphics[width=.9\linewidth]{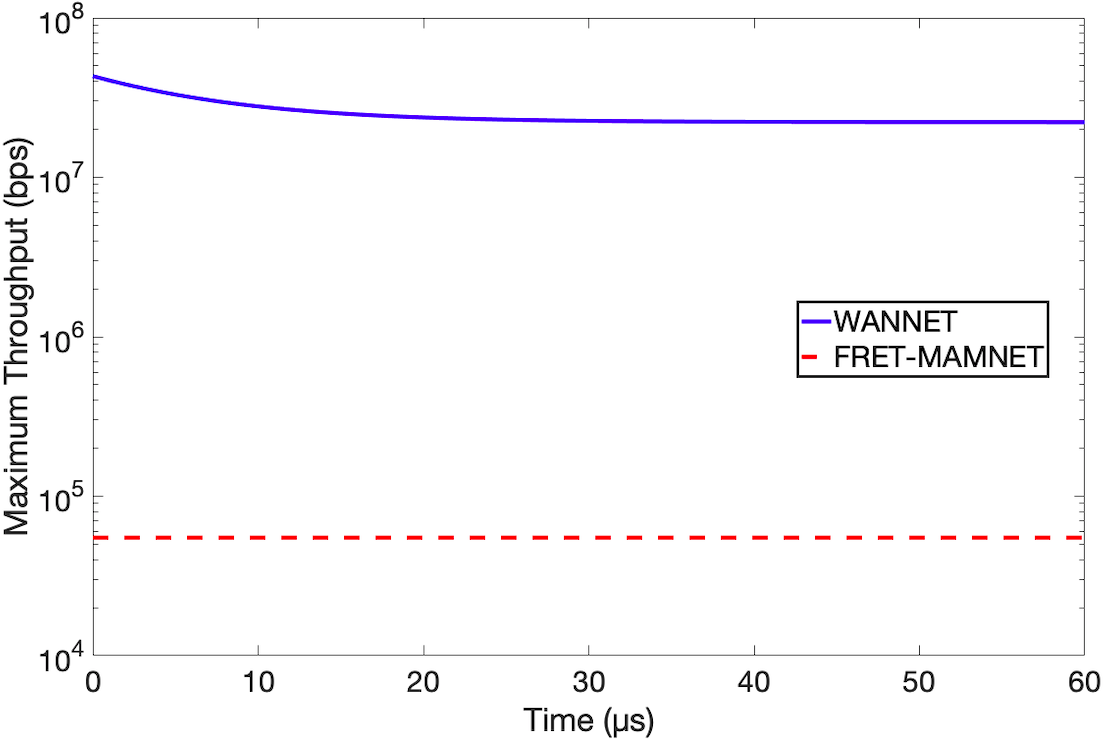}
		\caption{Maximum Throughput comparison between the proposed WANNET and FRET-MAMNET.}
		\label{fig:9}
	\end{figure} 
	
	In future work, we will study the MAC and application layers of the proposed WANNET, and we will propose a protocol for each layer. Furthermore, we will also propose a protocol for error correction by using DNA computation.

\end{document}